\documentclass[twocolumn]{aa}
\usepackage{graphicx}
\usepackage{amsmath}
\newcommand{\be}{\begin{equation}}
\newcommand{\ee}{\end{equation}}
\newcommand{\ba}{\begin{eqnarray}}
\newcommand{\ea}{\end{eqnarray}}
\newcommand\simgreater{\buildrel > \over \sim}

\begin{document}

\authorrunning{Geppert, K\"uker \& Page}

\titlerunning{Temperature distribution in magnetized neutron star crusts II}
	      
\title{Temperature distribution in magnetized neutron star crusts. II.\\
        The effect of a strong toroidal component}
              
\author{U. Geppert\inst{1}, M. K\"uker\inst{2} \and
Dany Page\inst{3}}
\institute{
Max-Planck-Institut f\"ur extraterrestrische Physik, Giessenbachstrasse, PF1312, D-85741 Garching\\
        \and
Astrophysikalisches Institut Potsdam, An der Sternwarte 16, D-14482 Potsdam, Germany \\
	\and
Instituto de Astronom\'{\i}a, UNAM, 04510 Mexico D.F., Mexico}
\offprints{U. Geppert, \email{urme@mpe.mpg.de}}
\date{}

\abstract{
We continue the study of the effects of a strong magnetic field on the temperature 
distribution in the crust of a magnetized neutron star (NS) and its impact on
the observable surface temperature. 
Extending the approach initiated in Geppert et al. (\cite{GKP04}), we consider more complex 
and, hence, more realistic, magnetic field structures but still restrict ourselves
to axisymmetric configurations.
We put special emphasis on the heat blanketing effect of a toroidal field component. 
We show that asymmetric temperature distributions can occur and a crustal field consisting 
of dipolar poloidal and toroidal components will cause one polar spot to be larger than the opposing one.
These two warm regions can be separated by an extended cold equatorial belt.
As an example we present an internal magnetic field structure which can explain,
assuming local blackbody emission, both the X-ray and optical spectra of the
isolated NS RXJ 1856-3754, the hot polar regions dominating the X-ray flux and
the equatorial belt contributing predominantly to the optical emission.
We investigate the effects of the resulting surface temperature profiles on 
the observable lightcurve which an isolated thermally emitting NS would produce for 
different field geometries. 
The lightcurves will be both qualitatively (deviations from sinusoidal shape) and 
quantitatively (larger pulsed fraction for the same observational geometry) 
different from those of a NS with an isothermal crust. 
This opens the possibility to determine the {\em internal} magnetic field strengths
and structures in NSs by modeling their X--ray lightcurves and spectra. The striking similarities of our model calculations with the observed spectra and pulse profiles of isolated thermally emitting NSs is an indication for the existence of strong magnetic
field components maintained by crustal currents.

}
\maketitle

\section{Introduction}

It has been understood since Greenstein \& Hartke (\cite{GH83}) that in presence of 
a sufficiently strong magnetic field, $\ge 10^{10}$ G, the surface temperature of
a neutron star (NS) will not be uniform as is expected in the unmagnetized case. 
Anisotropy of heat transport, caused by both classical and quantum magnetic field effects, 
in the thin low density ($\rho \le 10^{10}$ g cm$^{-3}$) upper layer, the so called 
{\em envelope}, results in a strongly reduced conductivity in the direction perpendicular 
to the field and an enhanced one along the field.
As a result, the regions around the magnetic poles are expected to be significantly warmer
than the regions around the magnetic equator.
Page  (\cite{P95}) and Page \& Sarmiento  (\cite{PS96}) applied the 
Greenstein \& Hartke formula to explain the lightcurves of the isolated thermally 
emitting NSs PSR 0833-45 (Vela), PSR 0656+14, PSR 0630+178 (Geminga) and PSR 1055-52,
considering dipolar field configurations, and also the addition of a quadrupolar component,
and including the General Relativistic curvature effects on photon trajectories.
Much work has been dedicated to study the effects of the magnetic field on
the properties of the NS envelope and crust (for reviews, see 
Yakovlev \& Kaminker \cite{YK94}, Ventura \& Potekhin \cite{VP01}, Lai \cite{L01}).
Due to the shallowness of the envelope, $\sim$ 100 m, heat transport can be
treated in the plane parallel approximation as a one dimensional problem in
which the heat flux is purely radial with a, locally uniform, magnetic field having
some arbitrary orientation with respect to the radial direction.
The most recent studies (Potekhin \& Yakovlev \cite{PY01}, henceforth `PY01, and 
Potekhin et al. \cite{PYCG03}) have included the best up to date transport coefficients
and equation of state: they showed some deviations from the simple Greenstein \& Hartke formula 
and finally give us a reasonably accurate description of a magnetized NS envelope.

In presence of a sufficiently strong magnetic field, $\simgreater 10^{12} - 10^{13}$ G,
the anisotropy of heat transport, simply due to the classical effect of Larmor rotation
of the electrons, extends to much higher densities and can even be present within the
whole crust.
Recently, we have shown (Geppert et al. \cite{GKP04}, subsequently Paper I) 
that in cases where the field geometry in the crust is such that the meridional component of
the field dominates over its radial component in a large part of the crust, as is the 
extreme case of a magnetic field entirely confined to the crust, the non uniformity of the
temperature, previously considered to be restricted to the envelope, may actually extend to the 
whole crust.
This modified crustal temperature results in a different surface temperature distribution
than what the simple Greenstein \& Hartke model predicted. 
In contradistinction, in case the dipolar poloidal field has its currents localized only in 
the core of the star, i.e., the field in the crust is considered as the one of a vacuum dipole,
the crust is practically isothermal and the non uniformity of the surface temperature is then
only due to field effects in the envelope.
This result, that the geometry of the magnetic field {\em in the interior} of the NS
leaves an observable imprint at the surface, potentially allows us to study the internal
structure of the magnetic field through modelling of the spectra and pulse profile of
thermally emitting NSs.

There exist growing observational evidence that the anisotropy of the heat 
transport in the envelope alone, assuming an otherwise isothermal crust,
can not explain the surface temperature distributions of some observed NSs.
The seven X-ray dim isolated NSs (XDINSs), dubbed "The Magnificent Seven"
(for reviews, see, e.g., Haberl \cite{H04,H04b,H05}) are nearby isolated
NSs, all discovered in the X-ray band where they show a thermal spectrum.
As a set they share several properties:
1) low X-ray absorptions imply small distances, $<$ 200 -500 pcs, 
2) high blackbody  temperatures deduced from the fit of their X-ray spectra indicate ages
of the order of $10^6$ yrs, possibly confirmed by measurements of proper motions which
allow for tentative identification of their birth sites in clouds of the Gould's belt,
3) no detection of radio emission,
4) no detection of a hard X-ray tail, typical of magnetospheric emission in active pulsars, 
5) no association with a supernova remnant,
6) detected spin periods (in 5 cases) between 3 to 11 s,
7) estimates of their surface dipolar field strength $B_0$ between $1 - 10 \times 10^{13}$ G,
either from a measurement of the period derivative $\dot{P}$ (1 case) or the interpretation of
broad absorption lines (in 5 cases) in the X-ray spectrum as being proton cyclotron lines,
8) optical broad band photometric detections (5 cases) can be interpreted as being the
Rayleigh-Jeans tail of a blackbody.
However, these optical data are well above the Rayleigh-Jeans tail of the
blackbody detected in the X-ray (``optical excess'') and indicate the presence of an
extended cold component of much larger area than the warm component observed in X-ray,
the latter having an emitting radius ($\sim 3-5$ km) much smaller that the usually assumed 
radius of a NS($\sim 10 - 15$ km).
Schwope et al. (\cite{SHHM05}) tried to fit the lightcurve of  RBS 1223 and concluded that
only a surface temperature profile with relatively small, about 4-5 kms across, 
hot polar regions may explain the observations. 
Pons et al. (\cite{Petal02}) and Tr\"umper et al. (\cite{TBHZ04}) arrived 
qualitatively at the same conclusion when they fitted the combined X-ray and
 optical spectrum of  RX J1856.5-3754.
In both cases, the smallness of the hot region is much below what can be reached
by considering anisotropic heat transport limited to only a thin envelope.
There are, however, not many possibilities to produce non uniform surface 
temperature distributions. 
Principially, rotation leads to an oblateness which enhances the thickness 
of the insolating outer layer at the equatorial belt making that region cooler 
than the poles (e.g. Geppert \& Wiebicke \cite{GW86}). 
This effect is, however, even for the fastest rotating millisecond pulsars, 
too small and a fortiori neglibigle for those isolated thermally emitting NSs whose 
rotational period is in the order of seconds. 
Another way to heat the polar regions is the bombardment by ultrarelativistic 
charged particles as can be expected in radiopulsars (see Gil et al. \cite{GMG03}).
Such heating is mostly confined to the polar caps whose angular radius can be roughly
estimated as $\theta_\mathrm{pc} \sim (2\pi R/cP)^{1/2}$, $R$ being the star's radius,
$c$ the speed of light and $P$ the pulsar's rotational period, giving a polar cap radius
$r_\mathrm{pc} \sim 0.46 \; P^{-1/2}$ km (for $P$ measured in sec.), which is much
smaller than the estimated size, a few kms, of the warm regions in the X-ray dim
isolated NSs whose periods are in the range of 3 to 11 s.

The previous Paper I  was devoted to demonstrate that a strong magnetic field
can have significant effects on the crustal temperature distribution.
We solved the stationary heat transport equation in Newtonian approximation 
and considered very simple, purely dipolar poloidal crustal field structures.
Our goal here is to consider more realistic models in which the currents maintaining
the poloidal magnetic field are distributed between the core and the crust and we also
consider the possible presence of a strong toroidal component in the crust.
We will still consider only axisymmetric, dipolar, magnetic field configuration.
Moreover, we now perform a wholly general relativistic formulation of the
heat transport and energy balance equations.

There exist also theoretical reasons that the presence of a strong toroidal
field is very likely. In a series of papers Markay \& Tayler \cite{MT73} (and
references therein) showed that a stable magnetic field configuration needs a
coexistence of poloidal and toroidal field components, having approximately
the same strength. Recently, Braithwaite \& Spruit (\cite{BS04}) considered the
stability of pure poloidal field geometries numerically in Ap stars and in
magnetic white dwarfs and arrived at the same conclusion.

Little is known about the magnetic field structure in NSs which is very
likely determined by processes during the proto-NS phase and/or in a 
relatively short period after that epoch. 
A proto-NS dynamo (Thompson \& Duncan \cite{TD93}) is unlikely 
to generate purely poloidal fields while differential  rotation will easily
wrap any poloidal field and generate strong toroidal components 
(Klu\'zniak \& Ruderman \cite{KR98}; Wheeler, Meier,  \& Wilson \cite{WMW02}).
The magneto-rotational instability (Balbus \& Hawley \cite{BH91}) also 
most certainly  acts in proto-NSs (Akiyama {\em et al.}  \cite{AWML03})
and results in toroidal fields from differential rotation (Balbus \& Hawley \cite{BH98}).
It is not yet explored whether the onset of supercondctivity in the core and the
relatively  fast crystallization of the largest part of the crust allows the field 
to relax into a stable (force free) state. 
However, it seems realistic to consider the effect of magnetic field 
configurations which consist of poloidal AND toroidal crustal components 
as well as of a star centered poloidal one. 
The relative strengths as well as the positions of the crustal components will 
depend on the - certainly varying - creation processes of those fields and 
will therefore be varied in reasonable limits. 

The paper is organized as follows: 
In the next section, \S~\ref{sec:magn-param}, we define the axisymmetric field structures 
and consider the resulting 2D heat transport and energy balance equations.
The next section, \S~\ref{sec:results1}, presents our results for the crustal temperature
distributions resulting from a large set of different field geometries and
\S~\ref{sec:results2} studies the observable effects of the resulting surface
temperature distributions, assuming simple isotropic blackbody emission.
As an example we present a crustal magnetic field structure which allows us to
reproduce the two temperature spectral fits of RX J1856.5-3754.
A discussion of our results and conclusions are the objects of \S~\ref{sec:discon}.

\section{Heat transport and magnetic field}
\label{sec:magn-param}


We will assume that the heat transport through the core
($\rho_{\mathrm{core}} \simgreater 1.6\cdot 10^{14}$g cm$^{-3}$) 
is practically unaffected by the magnetic field.
This assumption is justified by the very likely occurence of proton superconductivity
which, in the usually assumed case that it is a type II superconductor, will confine
the magnetic flux to fluxoids of microscopic size leaving the bulk of the core matter
devoid of magnetic field\footnote{In case the protons form a type I superconductor, 
as has been proposed by
Link (\cite{L03}) and Buckley, Metliski \& Zhitnisky (\cite{BMZ04}), this
assumption would have to be revised.}.
As a consequence we consider the stellar core to be isothermal and
we only need to define the magnetic field structure and calculate the temperature
distribution in the crust.
Moreover, we will still, in the present work, restrict ourselves to axisymmetric configurations.

\subsection{Strcture of the Magnetic Field}
\label{B_crust}

The assumption of axial symmetry suggests a decomposition of the NS magnetic field
in a poloidal and a toroidal component (we refer the reader to R\"adler \cite{R2000}
for a general presentation of the decomposition of magnetic fields into poloidal
and toroidal components) and we will restrict ourselves to the lowest order terms,
i.e., dipolar fields.
The boundary conditions at the NS surface are such that the internal field matches 
smoothly an external potential field; an observer above the NS surface would 
see only a dipolar poloidal field, whatever internal field structure is present.

Basically we will consider three components of the NS magnetic field:\\

1. The core field $\vec{B}^\mathrm{core}$ is a dipolar field having no 
currents in the crust and is maintained by axisymmetric currents in the core.
Then a vacuum-like dipolar field will penetrate the crust with components,
in spherical coordiinates $(r,\theta,\varphi)$ oriented along the symmetry axis,
\be
B_r^\mathrm{core}      =  B_0^\mathrm{core} \; \frac{\cos \theta}{x^3} \; ,
\ee
\be
B_\theta^\mathrm{core} =  - \frac{B_0^\mathrm{core}}{2} \; \frac{\sin \theta}{x^3} \; ,
\ee
\be
B_\varphi^\mathrm{core} = 0 \; ,
\ee
where $B_0^\mathrm{core}$ denotes the value of this component of the magnetic field
at the magnetic pole on the surface of the star and $x=r/R_{N\!S}$ 
the dimensionless radial coordinate, normalized on the NS radius $R_{N\!S}$.
This description of the core field in the crust assumes that it feels the 
crustal shell above the crust-core boundary as vacuum.

2. The crustal poloidal magnetic field $\vec{B}^\mathrm{crust}$ is maintained
by toroidal currents circulating in the crust. As for the core field
we will consider here only its largest scale component, which continues
beyond the NS surface as a dipolar vacuum field and can be 
conveniently described in terms of the (possibly time dependent) Stokes stream 
function $S = S(r,t)$.
 
The field components are then expressed as
\be
B_r^\mathrm{crust}      = \frac{2 \cos \theta}{r^2} S(r,t) 
         = B_0^\mathrm{crust} \; \frac{\cos \theta}{x^2} \; s(x,t)
\label{equ:Stokes-r}
\ee
\be
B_\theta^\mathrm{crust} =  \frac{\sin \theta}{r}
             \frac{\partial S(r,t)}{\partial r}
         =  \frac{B_0^\mathrm{crust}}{2} \; \frac{\sin \theta}{x}
             \frac{\partial s(x,t)}{\partial x}
\label{equ:Stokes-theta} 
\ee
\be
B_\varphi^\mathrm{crust} = 0
\ee

where $B_0^\mathrm{crust}$ is - as for $B_0^\mathrm{core}$ - the strength of this
field component at the magnetic pole. 
We have introduced the normalized function $s=2S/B_0/R_{N\!S}^2$, which must satisfy the
boundary conditions:  $s=0$ at the crust-core boundary and $s=1$ with
$\partial s/\partial x = -1$ at the surface.

3. The crustal toroidal magnetic field $\vec{B}^\mathrm{tor}$ is maintained
by poloidal currents circulating in the crust.
In an axisymmetric configuration it vanishes at the NS surface and we will assume
it does not penetrate the core.

In these condition $\vec{B}^\mathrm{tor}$ has vanishing components in the $r$ and $\theta$
directions and, applying the same formalism as led to the representation of
the poloidal field by the scalar quantity $s$ (see e.g.: R\"adler \& Geppert  \cite{RG99}), 
we can express the dipolar component of the toroidal field via the scalar $t(r,t)$ by
\be 
B^\mathrm{tor}_{\varphi}= \frac{\sin \theta}{r} T(r,t) = 
                        B^\mathrm{tor}_0 \frac{\sin \theta}{x} t(x,t)
\label{equ:Stokes-tor} 
\ee
and $t$ must vanish at the surface and at the crust-core boundary while its maximum
value within the crust is fixed to 1.
Notice that this component is symmetric with respect to the equatorial plane while
the dipolar poloidal fields are antisymmetric and, hence, the total magnetic field is
asymmetric.
To maintain an antisymmetric field configuration, to a dipolar poloidal field one would have
to add a quadrupolar toroidal  $\vec{B}^\mathrm{tor} \propto \sin{\theta}\cos{\theta}$,
with all toroidal components with odd multipolarity vanishing.

Determination of the spin-down of a NS can give us an estimate of the
strength of the dipolar poloidal field at the magnetic pole, $B_0$.
In our models $B_0 = B^\mathrm{crust}_0 + B^\mathrm{core}_0$: this gives us no information 
about either the possible values of $B^\mathrm{tor}$, and the relative strengths of
$B^\mathrm{core}_0$ and $B^\mathrm{crust}_0$, or about the shape of the two functions 
$s(x)$ and $t(x)$. 
There is thus a large degree of arbitrariness in the choice of the field structure, even
within our restriction to the dipolar terms, and we will consequently consider many
differente cases.
We choose for $s(r)$ one of the models of Page et al. (\cite{PGZ00})
at an age $t \simeq 10^5$ yrs: in these time evolution models of crustal field,
the field was not allowed to penetrate the stellar core due to proton superconductivity
and the crustal currents sustaining the field were slowly migrating toward the 
high conductivity layers near the crust-core boundary. 
This gives us some confidence that our choice of $s(x)$ is not too unrealistic\footnote{Notice,
however, that the models of Page et al. (\cite{PGZ00}) used a 1D numerical code and hence
no $\theta$ dependance of the temperature in the crust, which is in contradiction with
the results of \S~\ref{sec:results1}.}.
For the function $t(x)$ we use a Gaussian-like shape with a maximum at
\be
x_\mathrm{max} = x_\mathrm{core} + \alpha \Delta x_\mathrm{crust}
\label{equ:xmax}
\ee
where $x_\mathrm{core}$ and $\Delta x_\mathrm{crust}$ are the normalized core radius and
crust thickness, resp., and consider three values of $\alpha$: 0.2, 0.5, and 0.7.
The normalized Stokes function $s(x)$ and the three different $t(x)$ functions we will
consider are plotted in Fig. 1 and the crustal field lines of the fields
are presented in the panels of Fig. 2.
It is important to notice that the maximum values of $B^\mathrm{core}$ is 
$B_0^\mathrm{core}/x_\mathrm{crust}^3 \simeq 1.3 B_0^\mathrm{core}$,
reached at the crust-core boundary, the maximum value of $B^\mathrm{tor}$ is $B_0^\mathrm{tor}$,
in the crust at the point $x = x_\mathrm{max}$, while the
maximum value of $B^\mathrm{crust}$ is about 13 times $B_0^\mathrm{crust}$, from its
$\theta$ component (Eq.~\ref{equ:Stokes-theta}), due to the fact that $s(x)$ must decrease 
from 1 at the surface to 0 at the crust-core boundary and hence have a large derivative 
$\partial s/\partial x \sim \Delta s/\Delta x_\mathrm{crust}$ with 
$\Delta x_\mathrm{crust} \sim 0.1$, as shown in Fig. 1

\subsection{The two--dimensional heat transport}
\label{Flux_B}

The thermal evolution of the crust is determined by
the energy balance equation which has, in axial symmetry and taking 
into account general relativistic effects, the following form:
\ba 
\frac{e^{-\Lambda}}{r^2} \frac{\partial}{\partial r}
(r^2~F_r~e^{2\Phi}) + \frac{e^{2\Phi}}{r~\sin \theta}
\frac{\partial}{\partial \theta} (\sin \theta F_{\theta}) =
\nonumber \\
 \left(e^{\Phi} C_v \frac{dT}{dt} + e^{2\Phi} Q_\nu \right)\, \,,
\label{equ:heat-balance}
\ea
where $T$ is the local temperature, $e^{\Phi},e^{\Lambda}$ 
are the redshift and length correction factor, $F_r$ and $F_{\theta}$ 
are the local radial and meridional components of the heat flux and $r$ and $\theta$ the local coordinates.
$Q_{\nu}$ and $C_v$ are the neutrino emissivity and specific heat, respectively, per unit volume.
We will study stationary configurations and neglect neutrino energy losses, i.e.,
replace the right-hand side of Eq.~\ref{equ:heat-balance} by zero, and rewrite this equation as
\ba
\frac{1}{x^2} \frac{\partial}{\partial \tilde{x}}
(x^2~\tilde{F}_r) + \frac{1}{x~\sin \theta}
\frac{\partial}{\partial \theta} (\sin \theta \tilde{F}_{\theta}) = 0
\label{equ:heat-balance2}
\ea
where $\partial/\partial\tilde{x} \equiv e^{-\Lambda} \partial/\partial x$ and
$\tilde{F}_{r,\theta} \equiv e^{2\Phi} F_{r,\theta}/R_{NS}$

In regions below the envelope, i.e. in the crust, the quantized motion of electrons 
transverse to the magnetic field lines doesn't play any role for the magnetic 
modification of the  heat transport. 
The magnetic field acts via the Larmor rotation of the electrons, the relevant 
heat carriers, impeding on the heat transfer perpendicular to the crustal field.
Introducing, as in Paper I, the components of the 
temperature gradient parallel and perpendicular to the magnetic field 
as well as the corresponding components of the heat conductivity 
tensor we arrive at the heat flux vector 
\ba
e^{\Phi}\vec{F} =
- \hat{\kappa} \cdot \vec{\nabla} (e^{\Phi}T) =
- \; \frac{\kappa_0}{1+(\omega_{\scriptscriptstyle B}\tau)^2} \times
\nonumber \\
        \left[ \vec{\nabla} (e^{\Phi}T) +
       (\omega_{\scriptscriptstyle B}\tau)^2 \; \vec{b} \; (\vec{\nabla} (e^{\Phi}T) \cdot \vec{b}) +
			\omega_{\scriptscriptstyle B}\tau \; \vec{b} \times \vec{\nabla} (e^{\Phi}T)
        \right]\,\, .
\ea
whose radial and meridional components are 
\ba
\tilde{F}_r & = & -\tilde{\chi}_1\frac{\partial \tilde{T}}{\partial \tilde{x}}
\nonumber \\
&+& \tilde{\chi}_2 \left(
    \frac{\partial \tilde{T}}{\partial \theta} \frac{s}{2x^4}
    \frac{\partial s}{\partial \tilde{x}}
    \sin{\theta}\cos{\theta}  - 
    \frac{\partial \tilde{T}}{\partial \tilde{x}}
    \frac{s^2}{x^4} \cos^2{\theta} \right)
\nonumber \\
&+&    \tilde{\chi}_3 \frac{t}{x^2}\frac{\partial \tilde{T}}{\partial \theta} \sin{\theta} \,\,,
\label{equ:F_r-chi}
\ea
\ba
\tilde{F}_{\theta} & = & -\tilde{\chi}_1\frac{1}{x} \frac{\partial \tilde{T}}{\partial \theta}
\nonumber \\
&+&    \tilde{\chi}_2 \left(
      \frac{\partial \tilde{T}}{\partial \tilde{x}} \frac{s}{2x^3}
      \frac{\partial s}{\partial \tilde{x}}
      \sin{\theta} \cos{\theta}  -
      \frac{\partial \tilde{T}}{\partial\theta}
      \frac{1}{4x^3}
      \left(\frac{\partial s}{\partial \tilde{x}}\right)^2
      \sin^2{\theta}\right)
\nonumber \\
&-&   \tilde{\chi}_3 \frac{t}{x}\frac{\partial \tilde{T}}{\partial \tilde{x}} \sin{\theta}   \,.
\label{equ:F_t-chi}
\ea
where we have written $\tilde{T} \equiv e^\Phi T$ and $\tilde{\chi}_i \equiv e^\Phi \chi_i$.

The heat conductivity coefficients are:
\be
\chi_1= \frac{\kappa_0}{1+(\omega_{\scriptscriptstyle B}\tau)^2}
\,\,,\,\,
\chi_2= \frac{\kappa_0\,(\omega_{\scriptscriptstyle B^p_0}\tau)^2}
        {1+(\omega_{\scriptscriptstyle B}\tau)^2}
\,\,,\,\,
\chi_3=\frac{\kappa_0\,
        (\omega_{\scriptscriptstyle B^t_0}\tau)}
        {1+(\omega_{\scriptscriptstyle B}\tau)^2})
\, ,
\label{equ:chi}
\ee
Note that the Larmor-frequencies $\omega_{\scriptscriptstyle B^{p,t}_0}$ in 
Eq.~\ref{equ:chi} are calculated with 
$B^p_0=B^\mathrm{core}_0 + B^\mathrm{crust}_0$ in $\chi_2$ and with
$B^t_0 = B^\mathrm{tor}_0$
for the ``toroidal'' heat conductivity coefficient $\chi_3$.
The Larmor-frequency in the denominator of these coefficients is calculated by using
the exact field strength at each point: 
$B^2=B^\mathrm{core}(r,\theta)^2 + B^\mathrm{tor}(r,\theta)^2 +B^\mathrm{crust}(r,\theta)^2$.

For a given magnetic field structure equation~\ref{equ:heat-balance} is solved with the heat flux components of equations~\ref{equ:F_r-chi} and \ref{equ:F_t-chi} numerically using the same code which has been briefly described in Paper I until a stationary solution is found. The neutrino emissivity in the crust can be neglected for isolated NSs of that age. The inner boundary condition, at the crust-core boundary is
$T(x=x_\mathrm{core},\theta) = T_\mathrm{core}$, i.e. the core is assumed to be isothermal.
For the outer boundary condition the argumentation given in section 2.4. of 
Paper I remains valid and we proceed in the same way:
we stop our integration at a density $\rho_\mathrm{b} = 10^{10}$ g cm$^{-3}$
and, at the corresponding $x_\mathrm{b}$ and for each $\theta$, the radial flux is given by
$F_r(x_\mathrm{b},\theta) = \sigma_\mathrm{SB} T_\mathrm{s}^4$
where $T_\mathrm{s}$ is the surface (effective) temperature corresponding to the
boundary temperature $T(x_\mathrm{b},\theta)$ obtained with the ``$T_\mathrm{b}-T_\mathrm{s}$''
resulting from the magnetized envelope models of PY01.

We use the same input physics as in Paper I for calculating the electron relaxation 
time $\tau$ and the same chemical composition of the crust.
For the core of the star we here used the recent models of 
Akmal, Pandharipande \& Ravenhall (\cite{APR98}).
We consider only a 1.4 M$_\odot$ NS, with a radius $R_{NS} = 11.4$ km,
$R^\infty \equiv R_{NS} \cdot e^{-\Phi}$ = 14.2 km, and
$x_\mathrm{crust} = 0.914$, $\Delta x_\mathrm{crust} = 0.086$.

The essence of the effect of a strong magnetic field is that the heat flux $\vec{F}$
is forced to be almost aligned with the local field $\vec{B}$ when 
$(\omega_{\scriptscriptstyle B}\tau)^2 \gg 1$ since then the component of the thermal conductivity 
tensor $\hat{\kappa}$ parallel to $\vec{B}$ is $\kappa_\| = \kappa_0$ while in the perpendicular
directions it is 
$\kappa_\perp = \kappa_0/(1+(\omega_{\scriptscriptstyle B}\tau)^2) \ll \kappa_\|$.
Values of $\omega_{\scriptscriptstyle B}\tau$ were plotted in Fig 2 of Paper I.
Notice that we do not calculate $F_{\varphi}$.
By axial symmetry this component is independent of $\varphi$ but certainly {\em not} equal to zero,
in spite of having $\partial T/\partial \varphi \equiv 0$\footnote{This is 
a strong limitation of our simplifying assumption of axial symmetry and can only be 
removed by using a 3D code which, hopefully, will be studied in the near future.}.
Since, for strong fields, heat essentially flows along the field lines, when
$\vec{B}^\mathrm{tor}$ is dominant $F_{\varphi}$ will also be much larger than $F_{\theta}$ 
and $F_r$ and produce a winding of the heat flow around the symmetry axis:
$\vec{F}$ follows the shortest possible paths with the highest possible conductivity and 
this winding effectively acts as a heat blanket.
\section{Crustal temperature profiles}
\label{sec:results1}

In this section we will examine the temperature distributions resulting from a set of 
representative magnetic field configurations in which the three dipolar components
$\vec{B}^\mathrm{crust}$, $\vec{B}^\mathrm{core}$, and $\vec{B}^\mathrm{tor}$ 
have different relative stengths.
We are faced with many parameters since beside the strengths of each component we
have also the essentially unknown locations of the crustal currents sustaining
the crustal fields, i.e., the two functions $s$ and $t$ of Eq.~\ref{equ:Stokes-theta},
\ref{equ:Stokes-r}, and \ref{equ:Stokes-tor}.
The cases of purely poloidal fields, i.e., $\vec{B}^\mathrm{tor} \equiv 0$, with only
$\vec{B}^\mathrm{crust}$ or $\vec{B}^\mathrm{core}$, but not both components
superposed, were studied in Paper I and we will describe here a large set of models
which, we hope, will allow us to get a good understanding of the joint effects
of these, still simplified axisymmetric dipolar, multicomponent field configurations.

We consider first the case where the whole magnetic flux is produced in the crust and
the field does not penetrate the core, i.e.,  
$\vec{B} = \vec{B}^\mathrm{crust} + \vec{B}^\mathrm{tor}$ and $\vec{B}^\mathrm{core} \equiv 0$.
Fig. 3 and 4 show the crustal temperature distributions,
and the resulting surface temperatures, for two different locations of the toroidal components,
with $x_\mathrm{max} = 0.7$ and $0.2$, respectively, and in each case for three different 
strengths of $B^\mathrm{crust}_0$.
Panels a) and b) of Fig. 3 show that, given the high $T_\mathrm{core}$,
a $B^\mathrm{crust}_0$ of $10^{12}$ or $10^{13}$ G is not strong enough to affect the 
heat flow, in agreement with our previous results of Paper I:
the inner part of the crust is essentially isothermal and it is the strong 
$\vec{B}^\mathrm{tor}$ which act as a heat barrier, through the winding of the field lines
in the $\varphi$ direction.
With a very strong $\vec{B}^\mathrm{crust}$ however, as in panel c) of 
Fig. 3, the thermal insulation is then practically entirely produced
by this component and $\vec{B}^\mathrm{tor}$ is located too far out to play any noticable role.
In Fig. 4 we illustrate the same cases of $\vec{B}^\mathrm{crust}$
but with $\vec{B}^\mathrm{tor}$ now located at higher densities: 
for $B^\mathrm{crust}_0 = 10^{12}$ and $10^{13}$ G, panels a) and b), we see again that the
thermal insulation is provided by the toroidal component but with the new effect that heat
can flow ``around'' the toroidal field, from the magnetic axis toward the equator, in the
outer part of the crust and result in much smaller meridional temperature gradients in this
region.
(This flow of heat ``around'' the toroidal field can also be seen in Fig. 3
but is much less efficient.)
The case of a very strong $\vec{B}^\mathrm{crust}$, panel c) of Fig. 4,
with this inner toroidal component is almost identical to the similar case of panel c)
of Fig. 3.

In the next two Figures 5 and 6
we explore the effect of the distribution of the poloidal flux between the crust and
the core, together with a toroidal component, i.e., we consider the three components
$\vec{B}^\mathrm{crust}$, $\vec{B}^\mathrm{core}$, and $\vec{B}^\mathrm{tor}$ together.
We restrict ourselves to a $\vec{B}^\mathrm{tor}$ whose maximum
is located deep inside the crust, i.e., $\alpha = 0.2$ or $0.5$ in Eq.~\ref{equ:xmax},
which, as seen above, is not as effective as when 
located further out but is possibly more realistic since one expects currents to
migrate toward the highest electrical conductivity layers (Page et al. \cite{PGZ00}).
In panels a) and b) of Fig. 5 the crust and the core contribute
equally to the poloidal field and we see that when it is much weaker than the toroidal
component, panel a), the latter acts as a heat barrier while in panel b) the poloidal
component is sufficiently strong to allow heat propagation in the radial direction, superposed
of course to the winding flow in the $\varphi$ direction, along the field lines
and only the region within about 30$^\circ$ above and below the equator where the 
field lines are not connected to the core are slightly colder.
In case the poloidal flux comes almost entirely from the core, 
Fig. 5 panel d), the insulating effect of the toroidal field
almost completely disappear.
Assuming equipartition of the poloidal flux between the core and crust components, 
i.e. $\sim$ 10\% coming from the crustal comonent, as in 
Fig. 5 panel c), the result is very similar to panel c) despite of
the fact that in this case the maximum of $\vec{B}^\mathrm{crust}$ is about
$1.3 \times 10^{13}$ G (from its $\theta$ component, see Fig. 1), i.e.,
larger than $\vec{B}^\mathrm{core}$.

We consider now an even stronger $\vec{B}^\mathrm{tor}$ in Fig. 6
and see that its insulating effect is again very strong, due to the presence of the
$(\omega_{\scriptscriptstyle B}\tau)^2$ in the denominators of Eq.~\ref{equ:chi} which 
is now increased by almost one order of magnitude.
The results are perfectly in line with our description of the other previous cases:
1) the two models where the poloidal flux is crust dominated exhibit larger insulating effects 
than the corresponging models where the flux is core dominated, see panels a) vs. c) and
panels b) vs. d), and
2) models where the toroidal field is located deeper inside the crust are less insulating than
when located further out, see panels b) vs. a) and panels d) vs. c), since in the former cases
heat can more easily flow arround the toroidal component back from the symmetry axis
toward the equator.

A noticeable general feature of all our results is the asymmetry between the two magnetic
hemispheres, resulting from the asymmetry of the total field $\vec{B}$, since
$\vec{B}^\mathrm{crust}$ and $\vec{B}^\mathrm{core}$ are anti-symmetric while
$\vec{B}^\mathrm{tor}$ is symmetric.
In all cases where a non-isothermality is apparent the warm region close to the symmetry
axis is broader in the lower hemisphere than in the upper one.
In cases where the poloidal component is almost comparable to the toroidal one, 
as in panels c)'s of Fig. 3 and 4, the asymmetry is
barely distinguishible but in all other cases it is clearly visible.

The lower panels of the four Figures 3 to 6
show the resulting surface temperature profiles $T(\theta)$, 
$\theta$ being the colatitude measured
from the symmetry axis, scaled to the temperature at the magnetic pole $\theta = 0$,
and compared, as dotted lines, to the naive $\cos^{1/2}(\Theta_B)$ scaling 
(which itself is a simplification of the Greenstein \& Hartke, \cite{GH83}, result).
These allow to easily produce the temperature distribution over the whole NS
surface and generate composite spectra and pulse profiles which we describe in the
next section.

\section{Observational consequences for thermally emitting neutron stars}
\label{sec:results2}

The very distinct surface temperature distributions resulting from non-isothermal crusts
as calculated in the previous section, and displayed in the lower panels of Figures 3 to 6, have several immediate obervational
consequences which we explore in this section.
In all our simulations of spectra and pulse profiles we use simple isotropic 
blackbody emission at each point of the stellar surface, with the local temperature
$T(\theta,\varphi)$, and include general relativistic light bending as described in
Page (\cite{P95}).

In presence of a strong toroidal field in the crust, the channeling of heat toward 
the two polar regions results in the appearance of two hot spots of very reduced
size. 
The Figure 7 shows five examples of surface temperature distributions
and the resulting observable pulse profiles. 
Naturally, models with the smallest hot spots result in the highest pulsed fractions, $P_f$,
with values above 30\%,
in contradistinction to the case of an isothermal crust which results in $P_f \sim 5$ \%.

The composite blackbody spectra resulting from the same five cases of Figure 7
are shown in Figure~\ref{fig:Spectra-Fig5}.
Assumed distance to the stars have been adjusted to give the same maximum flux in the
X-ray band, and thus very similar X-ray spectra:
given this adjustment the difference between the relative areas of the hot and cold regions 
in the various cases result in differences in the predicted optical fluxes.
Comparison, by eye, of the surface temperature plots, left panels of Figure 7,
with the relative optical fluxes shows a direct correlation between the relative size of the cold
region with the optical flux.
This is confirmed by Figure 9 in which the percentage of the total surface area
whose temperature is below a certain value is shown and exhibit the same ordering as the
optical fluxes of Figure 8.
This correlation can easily be made quantitative by noticing that most of the flux comes into X-ray and the latter is thus given by $F_X \simeq a (R/D)^2 \sigma_\mathrm{SB} T_\mathrm{eff}^4$, 
$D$ being the star's distance and $a$ a correction factor for the interstellar absorption, 
where the effective temperature $T_\mathrm{eff}$ is naturally defined by
\be
T_\mathrm{eff} \equiv 
\left[\int \!\!\! \int \frac{d\Omega}{4\pi} \; T_s(\theta,\varphi)^4 \right]^{1/4}
\label{equ:Teff}
\ee
The optical flux, however, corresponding to the Rayleigh-Jeans tail of the spectrum is given by
$F_O \simeq (R/D)^2 A T_\mathrm{ave}$, $A$ being a constant dependent on the energy range
included in the "optical", and the average temperature $T_\mathrm{ave}$ being also naturally defined as
\be
T_\mathrm{ave} \equiv \int \!\!\! \int \frac{d\Omega}{4\pi} \; T_s(\theta,\varphi)
\label{equ:Tave}
\ee
One hence has the simple relationship
\be
F_O \simeq  F_X  \frac{T_\mathrm{ave}}{T_\mathrm{eff}^4}  \; \times \; \frac{A}{a \; \sigma_\mathrm{SB}}
\label{equ:F_O}
\ee
The spectra plotted in Figure 8 considered different distances $D$, adjusted
to  produce almost identical X-ray flux but with the same $N_H$ and hence approximately the 
same $a$'s, and Eq.~\ref{equ:F_O} allows to correctly predict the relative $F_O$'s, within 10\%,
using the values of ($T_\mathrm{ave}$,$T_\mathrm{eff}$) given in Figure 7.

As a concrete example, we show in Figure 10 an attempt to model the
spectrum of the isolated NS RX J1856.5-3754.
Its X-ray spectrum is well fitted by a hot blackbody, with $T^\infty \sim 6 -7 \times 10^5$ K and
an effective radius $R_\mathrm{H}^\infty \sim 4.5$ km, 
while the optical data can be interpreted as the Rayleigh-Jeans tail of a cold blackbody
component with $T^\infty \leq 3.5 \times 10^5$ K and an effective radius 
$R_\mathrm{C}^\infty \geq 17$ km,
assuming a distance $D = 117$ pcs (Pons et al. \cite{Petal02}; Tr\"umper et al. \cite{TBHZ04})
which are plotted in this figures as dashed lines.
Using the field structure of panel c) of Figure 6 and the resulting
surface temperature profile we obtain a reasonable fit, shown as a continuous line,
to this two blackbodies description.

\section{Discussion and conclusions}
\label{sec:discon}

We have studied the impact of different crustal magnetic field configurations on the 
transport of heat within the crust of a NS and the resulting surface temperature 
distributions with their observable consequences. 
In a previous paper (Paper I) we had first considered the simplest cases of a poloidal
dipolar field either entirely confined to the crust (``crustal field'') or exclusively
produced by currents in the core and hence having in the crust the same structure as
a vacuum dipole (``core field'').
In the present work we have considered the effect of an additional (dipolar) toroidal
component, located within the crust, with various combinations of core and crustal fields,
and all three components having the same symmetry axis.

Since we consider here only poloidal and toroidal dipolar components, due to 
the axial symmetry the toroidal field component is confined to the interior of the NS
and the external field geometry is always a purely poloidal dipolar one uniquely
characterized by its strength at the magnetic pole, $B_0$.
Therefore, the effects of various magnetic field structures with the same $B_0$ 
in the magnetosphere, as e.g. pulsar properties and rotational evolution of the NS, 
are indistinguishable.
However, these different inner geometries of the magnetic field result in very
different surface temperature distributions which are potentially distinguishable
through observations of thermal emission from isolated cooling NSs.

We had shown in Paper I that a strong magnetic field practically channels the heat flow
along its field lines.
We had found that a core field, with field lines essentially radial in most of the crust, 
results in an almost isothermal crust and the surface temperature is hence only controlled
by the anisotropic transport in the uppermost layers of the envelope
(Greenstein \& Hartke \cite{GH83}, Page \cite{P95}) exhibiting two symmetric extended warm
regions, one in each hemisphere, and a cold belt along the magnetic equator.
In contradistinction, a crustal field, having field lines with large meridional components,
can produce strongly non-isothermal crusts and the heat flow is much more concentrated around
the magnetic poles, resulting in a much wider cold equatorial belt.
Our main result here is that the addition of a toroidal component, inside the crust, is able
to add a very efficient heat blanket which may force the heat to flow within a narrow 
region along the polar axis and result in a cold region covering most of the stellar surface
with two warm spots arround the two magnetic poles.
However, we find that for this effect to occur it is essential to have a significant part of
the poloidal field produced in the crust: the toroidal field in itself is not sufficient
in case the field lines of the poloidal component are almost radial.

The presence of two small warm regions separated by an extended cold belt has two inmediate
observational consequences.
The first is that the observable pulsed fraction in the X-ray band can be very large, above
30\% assuming isotropic blackbody emission, and, second, the cold region's emission contributes
little to the X-ray flux but dominates the detectable flux in the optical range, appearing as 
an ``optical excess''.
Notice, however, that addition to the dipolar poloidal field of a quadrupolar component also
allows, potentially, to reach high pulsed fractions (Page \& Sarmiento \cite{PS96})
within the same isotropic blackbody emission scheme and without any todoidal component.
In this latter case, high $P_f$'s are reached when the quadrupole sufficiently deforms 
the field to push the two magnetic poles close to each other but without significantly
altering the star's effective and average temperature 
(see, e.g., Figure 1 of Page \& Sarmiento \cite{PS96}), and hence producing no significant
``optical excess''.
The ``Magnificent Seven'' briefly presented in the Introduction seem to be very good
candidate NSs for which our results may be relevent.
Their estimated dipolar surface field strengths $B_0$ are above $10^{13}$ G, opening the
possibility of strong anisotropic heat transport in their crust, and the ``optical excess''
observed in 5 cases is then a natural consequence of the large $B_0$'s in case
their crusts harbor very strong poloidal fields.
As we has seen, channeling of the heat flux toward two small polar regions requieres
moreover that part of the poloidal flux be confined to the crust:
this can be achieved either by the action of some dynamo process which produced such
a poloidal field component or/and by the expulsion of flux from the core through the
action of the outward migration of the neutron superfluid fluxoid resulting from the
NS spin-down and the consecutive expulsion of the magnetic fluxoid resulting
from the core proton superconductor.
This latter process is likely to have occured in the ``Magnificent Seven'' given
their unusually long period, for their relatively young ages, which are very likely 
the results of fast spin-down due to their strong dipolar poloidal fields.
However, our axisymmetric field configurations produce symmetric, but not sinusoidal,
light-curves and the superposition of a quadrupolar component may be necessary to
produce precise fits of the observed pulse profiles of the seven XDINS 
(Zane \& Turolla \cite{ZT05})

Results very similar to the ones presented here have been recently obtained by
P\'erez-Azor\'in, Miralles \& Pons (\cite{PAMP05}) who also conclude that a toroidal
component localized within the crust lead to surface temperature distributions
with very localized warm regions and an extended cold equatorial belt.
These authors also took into account the transport of heat by phonons, which is
unaffected by the magnetic field and thus reduces the anisotropicity, and the small
quantum corrections present at high densities for very strong fields, two effects
we did not consider.
Moreover, they performed their heat transport calculations to lower densities than
we have considered here and assumed that the upper layer is cut at a finite density
due to solidification of the material while we simply assumed that the envelope extends
into an atmosphere and ``glued'' our models with pre-existent envelope models.
Nevertheless, despite of these differences, the similarity of the results prove
that the effect of a toroidal field is independent of the details and make our
common results very robusts.

Claiming that the observed optical-X-ray properties of the ``Magnificent Seven''
prove the existence of strong toroidal fields in their crust may still be premature.
Given our poor understanding of the emissive properties of the NS surface,
being it either an atmosphere, a liquid or a solid depending on its chemical
composition, temperature and magnetic field strength, one cannot exclude
that the ``optical excess'' is achieved locally, i.e., is an intrinsic feature of 
the spectrum produced at each point of the surface, instead of being a global effect
due to strong surface temperature differences as we have obtained.
In contradistinction, three very active pulsars exhibit very different properties
(Vela: Romani, Kargaltsev \& Pavlov 2005, \cite{RKP05}, 
PSR B0656+14: Shibanov et al. \cite{SSLGL05},
and Geminga: Kargaltsev et al. \cite{KPZR05}): a thermal spectrum is clearly
seen in the X-ray but near \& far UV observations give upper limits on its Rayleigh-Jeans
tail which are in agreement or even {\em below} the extrapolations of the X-ray spectrum.
The reasons for this different behavior is presently unknown and may be due to their weaker
magnetic field, different surface chemical composition, and/or reprocessing of the
thermal photons in the magnetosphere.

Finally, an issue raised by our models is the time evolution and stability of the
field configurations we have considered.
We intend to perform coupled time-dependent cooling and field evolution calculations
to try to eliminate at least the most unlikely configurations, i.e., either highly unstable
configurations and/or configurations based on current distribution which will evolve
fast and reajust or decay.
Nevertheless we can speculate that a very strong crustal toroidal field could produce such
strong tension that the crust should readjust itself, as is seen in magnetars, and
this may be the cause of the observed change in the spectrum and pulse profile of
RX J0720.4-3125 (de Vries et al. \cite{dVVMV04}).

\begin{acknowledgements}

Part of this work is supported by a binational grant from DGF-Conacyt
\#444MEX113/4/0-2.
D.P.'s work is partially supported by grants from
UNAM-DGAPA (\#IN112502) and Conacyt (\#36632-E).
\end{acknowledgements}
\section{Figure captions}

{\bf Fig. 1:} Continuous lines: normalized Stokes function $s(x)$ and its derivative
(scaled by a factor 10) used in this work, Eq.4 and 5. Discontinuous lines: the three different normalized functions $t(x)$ we consider for the toroidal field, Eq.7.

\noindent {\bf Fig. 2:} Qualitative sketch of the field geometries as considered here.
Left panel: field lines of the poloidal crustal field having outside the same
dipolar potential field as a star centered field (right panel). Middle panel: isolines of a toroidal field with its maximum in the middle of the crust ($\alpha = 0.5$).

\noindent {\bf Fig. 3:} Thermal structure of the crust of a neutron star.
(The radial scale of the crust is stretched by a factor of 5 for clarity.)
The magnetic field includes the $\vec{B}^\mathrm{crust}$ and $\vec{B}^\mathrm{tor}$ components and its structure is illustrated in panel d (field lines of the poloidal component and color coded isolines of the toroidal component normalized on $B^\mathrm{tor}_0$). The maximum of $B^\mathrm{tor}$ is located at
$x_\mathrm{max} = x_\mathrm{core} + 0.7 \, \Delta x_\mathrm{crust}$ (see Eq. 8).
Panels a,b, and c show the temperature distribution for  $B^\mathrm{tor}_0 = 10^{15}$G and $B^\mathrm{crust}_0 = 10^{12}, 10^{13}$, and $10^{14}$G, respectively.
The color coded temperatures are normalized on the core temperature,
$T_\mathrm{core}= 6\cdot 10^7$K.
The lower panels show the resulting surface temperature profiles $T(\theta)$,
once an extra insulating envelope has been glued to the core temperature distributions
of the upper panels, and the dotted lines illustrate the same profile when an
isothermal crust is assumed.

\noindent {\bf Fig. 4:} The same as in Fig. 3 but assuming a location of the maximum of the toroidal component much closer to the crust--core interface:
$x_\mathrm{max} = x_\mathrm{core} + 0.2 \, \Delta x_\mathrm{crust}$ (see Eq. 8)

\noindent {\bf Fig. 5:} The same as in Fig. 4 but now including the three components $\vec{B}^\mathrm{crust}$, $\vec{B}^\mathrm{core}$ and $\vec{B}^\mathrm{tor}$.
The magnetic field scales are in
panel a:     $B^\mathrm{crust}_0 = 10^{12}$G, $B^\mathrm{core}_0 = 10^{12}$G,
panel b:     $B^\mathrm{crust}_0 = 10^{13}$G, $B^\mathrm{core}_0 = 10^{13}$G,
panel c:     $B^\mathrm{crust}_0 = 10^{12}$G, $B^\mathrm{core}_0 = 10^{13}$G,
and panel d: $B^\mathrm{crust}_0 = 10^{11}$G, $B^\mathrm{core}_0 = 10^{13}$G.
In all four cases $B^\mathrm{tor}_0 = 10^{15}$ G and
$x_\mathrm{max} = x_\mathrm{core} + 0.2 \, \Delta x_\mathrm{crust}$ (see Eq. 8).

\noindent {\bf Fig. 6:} The same as Fig. 5 but assuming a stronger toroidal field with $B^{\mathrm{tor}}_0 = 3\cdot 10^{15}$G. In panels a and b the crustal poloidal field dominates the core field ($B^{\mathrm{crust}}_0 = 7.5\cdot 10^{12}$G, $B^{\mathrm{core}}_0 = 2.5\cdot 10^{12}$G) while in panels c and d the core field is dominant ($B^{\mathrm{crust}}_0 = 2.5\cdot 10^{12}$G, $B^{\mathrm{core}}_0 = 7.5\cdot 10^{12}$G). In panels a and c the toroidal field has its maximum in the middle of the crust, $x_\mathrm{max} = x_\mathrm{core} + 0.5 \, \Delta x_\mathrm{crust}$ (see Eq. 8),
while in panels b and d it is located closer to the crust-core boundary
$x_\mathrm{max} = x_\mathrm{core} + 0.2 \, \Delta x_\mathrm{crust}$.

\noindent {\bf Fig. 7:} Surface temperature distributions (left panels) in an area preserving
representation with a color scale following the emited flux ($\propto T^4$).
Panels a to d use the internal field structures of the corresponding panels in
Fig. 6 while panel e assumes an isothermal crust.
In all cases the dipolar symmetry axis is oriented in the rotational equatorial plane
and the right panels show the resulting pulse profiles (in arbitrary units) which
an observer, also located in the rotational equatorial plane, would detect.
In all cases the core temperature is the same but the star's distance has been
adjusted to give the same average flux (see Fig. 8).
Number pairs within parentheses on the left panels give
($T_\mathrm{ave}$,$T_\mathrm{eff}$), as defined in Eq. 15 and 16, resp., in units of $10^5$ K.
All 5 models have almost the same maximum surface temperature $T_\mathrm{max}
\simeq 8.45 \times 10^5$ K but different minimal temperatures $T_\mathrm{min}$.

\noindent {\bf Fig. 8:} Observable spectra for the five surface temperature distributions and
observable pulse profiles, labelled as ``a'' to ``e'', shown in Fig. 7.
The stars, with radii $R=11.4$ km and $R_\infty = 14.28$ km for a 1.4 $M_\odot$,
are assumed to be at distances of 100, 142, 131, 202, and 220 pc, resp.,
to produce almost identical observable spectra in the X-ray band
($N_H = 1 \times 10^{20}$ cm$^{-2}$ for interstellar absorption)
but resulting in significantly different fluxes in the optical range.

\noindent {\bf Fig. 9:} Distribution of the surface temperature in terms of the percentage of the superficial area whose temperature is lower than $T$, for the five models of Fig. 7 and 8

\noindent {\bf Fig. 10:} Fit of the spectrum of RX J1856.5-3754. Dotted lines show the two blackbodies fit to the data from Tr\"umper {\em et al.} (\cite{TBHZ04}).
The continuous line show our results: the star has a radius $R=14.4$ km and $R_\infty = 17.06$ km for a 1.4 $M_\odot$, at a distance of 122 pcs ($N_H = 1.6 \times 10^{20}$ cm$^{-2}$ for interstellar absorption) and the observer is assumed to be aligned with the rotation axis.
The magnetic field structure corresponds to model c of Figure 6
adjusted to the 14.4 km radius with $T_b = 6.8\times10^7$ K, resulting in
$T_\mathrm{eff}^\infty = 4.62\times10^5$ K and $T_\mathrm{max}^\infty = 8.54\times10^5$ K.



\end{document}